\documentclass[12pt]{article}
\usepackage{graphicx}

\def\re{\rangle}
\def\b{\begin{equation}}
\def\e{\end{equation}}

\title{Degradation of entanglement in moving frames}
\author{Shahpoor Moradi$^1$,
 \thanks{e-mail: shahpoor.moradi@gmail.com}}
\date{\today}
\begin{document}
\maketitle {\it \centerline{$^1$  Department of Physics, Razi
University, Kermanshah, IRAN}}
\begin{abstract}
The distillability of  bipartite entangled state  as seen by
moving observers has been investigated. It is found that the same
initial entanglement for a state parameter $\alpha$ and its
"normalized partner" $\sqrt{1-\alpha^2}$  will be degraded as seen
by moving observer. It is shown that in the ultra relativistic
limit, the state does not have distillable entanglement for any
$\alpha$.

\end{abstract}
Relationship between special relativity and quantum information
theory is discussed by many authors  \cite{p1}.
 Peres \textit{et
al} \cite{p2} investigated the relativistic properties of spin
entropy for a single, free particle of spin$-1/2$. They  show that
the usual definition of quantum entropy has no invariant meaning
in special relativity. The reason is that, under a Lorentz boost,
the spin undergoes a Wigner rotation  whose direction and
magnitude depend on the momentum of the particle. Even if the
initial state is a direct product of a function of momentum and a
function of spin, the transformed state is not a direct product.
Lamata \textit{et al}\cite{pa} define weak and strong criteria for
relativistic isoentangled and isodistillable states to
characterize relative and invariant behavior of entanglement and
distillability.  In this letter, we choose a generic state as the
initial entangled state and we will try to show that the
entanglement is degraded as seen by the relativistically observer.
This  help us to understand the relationship between special
relativity and quantum information theory. The initial nonmaximal
entangled state is \b
|\Phi\re=\alpha\Psi^{(a)}_1({\mathbf{p}_a})\Psi^{(b)}_1({\mathbf{p}_b})
+\sqrt{1-\alpha^2}\Psi^{(a)}_2({\mathbf{p}_a})\Psi^{(b)}_2({\mathbf{p}_b}),\e
where $\alpha$ is some number that satisfies $|\alpha|\in(0,1)$.
Here ${\mathbf{p}_a}$  and  ${\mathbf{p}_b}$ are the corresponding
momentums vectors of particles $A$ and $B$ and \b
\Psi^{(a)}_1({\mathbf{p}_a})=g({\mathbf{p}_a})|0\re=|0,{\mathbf{p}_a}\re=\left(%
\begin{array}{c}
  g({\mathbf{p}_a}) \\
 0 \\
\end{array}%
\right), \e \b
\Psi^{(a)}_2({\mathbf{p}_a})=g({\mathbf{p}_a})|1\re=|0,{\mathbf{p}_b}\re=\left(%
\begin{array}{c}
0 \\
 g({\mathbf{p}_a}) \\
\end{array}%
\right). \e For simplicity assume that they are sufficiently well
localized around momenta ${\mathbf{p}}$, Under the Lorentz
transformation the states (2) and (3) transformed as \cite{pa} \b
\Lambda[\Psi_1({\mathbf{p}})]=\left(%
\begin{array}{c}
  b_1({\mathbf{p}}) \\
  b_2({\mathbf{p}}) \\
\end{array}%
\right)=\left(%
\begin{array}{c}
  \cos(\theta_{{\mathbf{p}}}/2) \\
  \sin(\theta_{{\mathbf{p}}}/2) \\
\end{array}%
\right)g({\mathbf{p}})=\cos\theta_(\theta_{{\mathbf{p}}}/2)|0,{\mathbf{p}}\re+
\sin(\theta_{{\mathbf{p}}}/2)|1,{\mathbf{p}}\re, \e \b
\Lambda[\Psi_2({\mathbf{p}})]=\left(%
\begin{array}{c}
  -b_2({\mathbf{p}}) \\
  b_1({\mathbf{p}}) \\
\end{array}%
\right)=\left(%
\begin{array}{c}
 - \sin(\theta_{{\mathbf{p}}}/2)\\
  \cos(\theta_{{\mathbf{p}}}/2)\\
\end{array}%
\right)g({\mathbf{p}})=-\sin(\theta_{{\mathbf{p}}}/2)|0,{\mathbf{p}}\re+
\cos(\theta_{{\mathbf{p}}}/2)|1,{\mathbf{p}}\re, \e where
$\theta_{{\mathbf{p}}}$ is Wigner angle satisfies the relation \b
\tan\theta_{{\mathbf{p}}}=\frac{\sinh\xi\sinh\delta}{\cosh\xi+\cosh\delta}
,\e here $\cosh\xi=(1-\beta^2)^{-1/2}$ where  $\beta$ is boost
speed and  $\cosh\delta=p_0/m$.  Now under Lorentz transformation
the state transformed to (after tracing over momentum eigenket
$|{\mathbf{p}}\re$ ), $$ |\Phi\re_{\Lambda}=\left(\alpha
\cos^2(\theta_{{\mathbf{p}}}/2)+\sqrt{1-\alpha^2}\sin^2(\theta_{{\mathbf{p}}}/2)\right)|00\re+$$$$
\sin(\theta_{{\mathbf{p}}}/2)\cos(\theta_{{\mathbf{p}}}/2)(\alpha-\sqrt{1-\alpha^2})(|01\re+|10\re)+
$$\b\left(\alpha
\sin^2(\theta_{{\mathbf{p}}}/2)+\sqrt{1-\alpha^2}\cos^2(\theta_{{\mathbf{p}}}/2)\right)|11\re.
\e  A very popular measure for the quantification of bipartite
quantum correlations is the concurrence \cite{H}. This quantity
can be defined \b
C(\rho)=\max\{0,\lambda_1-\lambda_2-\lambda_3-\lambda_4\}. \e with
$\lambda_i$ being the square roots of the eigenvalues of
$\rho_{AB}(\sigma_y\otimes\sigma_y)\rho^*_{AB}(\sigma_y\otimes\sigma_y)$
where the asterisk denotes complex conjugation and and
$\sigma_y=\left(%
\begin{array}{cc}
  0 & -i \\
  i & 0 \\
\end{array}%
\right).$ Now the concurrence for this state is \b
C(|\Phi\re_{\Lambda})=2\alpha\sqrt{1-\alpha^2} .\e Which is the
Lorentz invariant concurrence. To be more precise one should take
wave packets in momentum space, with Gaussian momentum
distributions
$g({\mathbf{p}})=\pi^{-3/4}w^{-3/2}\exp\left(-|{\mathbf{p}}|^2/2w^2\right)$.
If we trace the momentum degrees of freedom we obtain the usual
entangled state $ |\phi\re=\alpha|00\re +\sqrt{1-\alpha^2}|11\re$.
 The general density matrix for two particle systems with
momentums ${\mathbf{p}}_a$ and ${\mathbf{p}}_b$ is given by \b
\rho_{\Phi}=\sum_{ijkl=1,2}C_{ijkl}\Psi_i({\mathbf{p}}_a)\otimes\Psi_j({\mathbf{p}}_b)
[\Psi_l({\mathbf{p}}'_a)\otimes\Psi_m({\mathbf{p}}'_b)]^{\dag}. \e
For state $(1)$ the coefficients $C_{ijkl}$ are
\[ C_{1111}=\alpha^2,\;\;\;\;C_{2222}=1-\alpha^2,\]\b
C_{1122}=C_{2211}=\alpha\sqrt{1-\alpha^2}
 \e
 For obtaining the Lorentz transformation of (10), we need the relativistic properties of spin
entropy for a single, free particle of spin$-1/2$. The quantum
state of a spin-$\frac{1}{2}$ particle can be written in the
momentum representation as follows
\b \Psi({\mathbf{p}})=\left(%
\begin{array}{c}
  a_1({\mathbf{p}}) \\
  a_2({\mathbf{p}}) \\
\end{array}%
\right), \e where
\b\int(|a_1({\mathbf{p}})|^2+|a_2({\mathbf{p}})|^2)d{\mathbf{p}}=1.\e
The density matrix corresponding to state $(6)$ is \b
\rho({\mathbf{p}}',{\mathbf{p}}'')=\left(%
\begin{array}{cc}
  a_1({\mathbf{p}}')a_1({\mathbf{p}}'')^* & a_1({\mathbf{p}}')a_2({\mathbf{p}}'')^* \\
  a_1({\mathbf{p}}')a_2({\mathbf{p}}'')^* & a_2({\mathbf{p}}')a_2({\mathbf{p}}'')^* \\
\end{array}%
\right).\e By setting  ${\mathbf{p}}'={\mathbf{p}}''={\mathbf{p}}$
and integrating over ${\mathbf{p}}$ we obtain the  reduced
density matrix for spin \b \sigma=\frac{1}{2}\left(%
\begin{array}{cc}
  1+n_z & n_x-in_y \\
  n_x+in_y & 1-n_z \\
\end{array}%
\right), \e where the Bloch vector $\mathbf{n}$ is given by \b
n_z=\int(|a_1({\mathbf{p}})|^2-|a_2({\mathbf{p}})|^2)d{\mathbf{p}}=1,\e
 \b n_x-in_y=\int  a_1({\mathbf{p}})
a_2({\mathbf{p}})^*d{\mathbf{p}}.\e  Now under Lorentz boost
density matrix (10) transformed into
$$\Lambda \rho_{\Phi} \Lambda
^{\dag}=\sum_{ijkl=1,2}C_{ijklmn}\Lambda
(p_a)\Psi_i({\mathbf{p}}_a)\otimes \Lambda
(p_b)\Psi_j({\mathbf{p}}_b)
$$\b\times[\Lambda (p_a)\Psi_l({\mathbf{p}}'_a)\otimes \Lambda (p_b)\Psi_m({\mathbf{p}}'_b)]^{\dag}. \e

The reduced density matrix for spin is obtained by setting
${\mathbf{p}}_a={\mathbf{p}}'_a$, ${\mathbf{p}}_b={\mathbf{p}}'_b$
and tracing over momentum
$$\tau=Tr_{{\mathbf{p}}_a{\mathbf{p}}_b}[\Lambda \rho_{\Phi}
\Lambda ^{\dag}]=\sum_{ijkl=1,2}C_{ijkl}Tr_{{\mathbf{p}}_a}
\left\{\Lambda ({{p}}_a)\Psi_i({\mathbf{p}}_a)[\Lambda
({{p}}_a)\Psi_k({\mathbf{p}}_a)]^{\dag}\right\}
$$\b\otimes Tr_{{\mathbf{p}}_b}
\left\{\Lambda ({{p}}_b)\Psi_j({\mathbf{p}}_b)[\Lambda
({{p}}_b)\Psi_l({\mathbf{p}}_b)]^{\dag}\right\}.\e
 To leading order $w/m\ll 1$ we have \b
n_z=n\approx1-\left(\frac{w}{2m}\tanh\frac{\xi}{2}\right)^2,\;\;\;\;n_x=n_y\approx
0,\e It can be appreciated in Eq. $(19)$ that the expression is
decomposable in the sum of the tensor products of $2\times 2$ spin
blocks, each corresponding to each particle. We compute now the
different blocks, corresponding to the four possible tensor
products of the states $(3)$ and $(4)$: \b Tr_{{\mathbf{p}}}
\left\{\Lambda (p)\Psi_1({\mathbf{p}})[\Lambda (p)\Psi_l({\mathbf{p}})]^{\dag}\right\}=\frac{1}{2}\left(%
\begin{array}{cc}
  1+n & 0 \\
  0 & 1-n \\
\end{array}%
\right), \e \b Tr_{{\mathbf{p}}}
\left\{\Lambda (p)\Psi_2({\mathbf{p}})[\Lambda (p)\Psi_2({\mathbf{p}})]^{\dag}\right\}=\frac{1}{2}\left(%
\begin{array}{cc}
  1-n & 0 \\
  0 & 1+n \\
\end{array}%
\right) ,\e \b Tr_{{\mathbf{p}}}
\left\{\Lambda (p)\Psi_1({\mathbf{p}})[\Lambda (p)\Psi_2({\mathbf{p}})]^{\dag}\right\}=\frac{1}{2}\left(%
\begin{array}{cc}
  0 & 1+n \\
  -(1-n) & 0 \\
\end{array}%
\right), \e \b Tr_{{\mathbf{p}}}
\left\{\Lambda (p)\Psi_2({\mathbf{p}})[\Lambda (p)\Psi_1({\mathbf{p}})]^{\dag}\right\}=\frac{1}{2}\left(%
\begin{array}{cc}
  0 &  -(1-n) \\
  1+n& 0 \\
\end{array}%
\right). \e With the help of Eqs . $(21)$-$(24)$, it is possible
to compute the effects of the Lorentz transformation, associated
with a boost in the $x$ direction, on any density matrix of two
spin-$1/2$ particles with factorized Gaussian momentum
distributions. In particular  density matrix (19) is reduced to
\b\tau=\frac{1}{4}\left(%
\begin{array}{cccc}
  4\alpha^2n+(1-n)^2 & 0 & 0 & 2\alpha\sqrt{1-\alpha^2}(1+n^2) \\
  0 & 1-n^2 & -2\alpha\sqrt{1-\alpha^2}(1-n^2) & 0 \\
  0 & -2\alpha\sqrt{1-\alpha^2}(1-n^2)  & 1-n^2 & 0 \\
  2\alpha\sqrt{1-\alpha^2}(1+n^2) & 0 & 0 & -4\alpha^2n+(1+n)^2 \\
\end{array}%
\right).\e We can apply now the positive partial transpose
criterion \cite{p3} to know whether this state is entangled and
distillable. The partial transpose criterion provides a sufficient
condition for the existence of entanglement in this case: if at
least one eigenvalue of the partial transpose is negative, the
density matrix is entangled; but a state with positive partial
transpose can still be entangled. It is the well-known bound or
nondistillable entanglement \cite{wid}. Partial transpose of
density matrix (25) yields
\[\tau^{T}=\frac{1}{4}\left(%
\begin{array}{cccc}
  4\alpha^2n+(1-n)^2 & 0 & 0 &-2\alpha\sqrt{1-\alpha^2}(1-n^2) \\
  0 & 1-n^2 & 2\alpha\sqrt{1-\alpha^2}(1+n^2) & 0 \\
  0 & 2\alpha\sqrt{1-\alpha^2}(1+n^2)  & 1-n^2 & 0 \\
  -2\alpha\sqrt{1-\alpha^2}(1-n^2) & 0 & 0 & -4\alpha^2n+(1+n)^2\\
\end{array}%
\right).\] It is possible diagonalize $\tau^{T}$  and get it's
eigenvalues \b \lambda_1
=\frac{1}{4}(1-n^2)+\frac{1}{2}\alpha\sqrt{1-\alpha^2}(1+n^2),\e
\b \lambda_2
=\frac{1}{4}(1-n^2)-\frac{1}{2}\alpha\sqrt{1-\alpha^2}(1+n^2),\e
\b \lambda_3
=\frac{1}{4}(1+n^2)+\frac{1}{2}\sqrt{n^2+\alpha^2(1-\alpha^2)(n^4-6n^2+1)},\e
\b \lambda_4
=\frac{1}{4}(1+n^2)-\frac{1}{2}\sqrt{n^2+\alpha^2(1-\alpha^2)(n^4-6n^2+1)}.\e
For $0<n,\alpha<1$ eigenvalues $\lambda_1$, $\lambda_3$ and
$\lambda_4$ are always positive. The eigenvalue $\lambda_2$ is
negative  for $\alpha\sqrt{1-\alpha^2}>R$ where \b
R=\frac{1-n^2}{2(1+n^2)}.\e In this range the logarithmic
negativity  takes the form
 \b
N=\log_2\left\{
\frac{1}{2}(1+n^2)\left(1+2\alpha\sqrt{1-\alpha^2}\right)\right\}.
\e In ultra relativistic limit $n\rightarrow 0$: $
N\rightarrow\log_2\left\{
\frac{1}{2}+\alpha\sqrt{1-\alpha^2}\right\}$, then  the state does
not have distillable entanglement for any $\alpha$.  For the rest
frame $n=1$: $N=\log_2\left\{ 1+2\alpha\sqrt{1-\alpha^2}\right\}$.
In the range $0<\alpha<1/\sqrt{2}$ the larger $\alpha$, the
stronger the initial entanglement; but in the range
$1/\sqrt{2}<\alpha<1$ the larger $\alpha$, the weaker the initial
entanglement. For finite velocity, the monotonic decrease of $N$
with increasing boost speed for different $\alpha$ means that the
entanglement of the initial state is lost due to Wigner rotation.
From Fig.1 it is found that the entanglement in moving frame, for
$\alpha$ and it's normalized partner $\sqrt{1-\alpha^2}$, will be
degraded as boost speed increases. Here we calculate the
concurrence which is defined as follows \b
C=2\sqrt{\det\rho_A}=2\sqrt{\det\rho_B} .\e For density matrix
(25) we have
\b \rho_A=\rho_B=\left(%
\begin{array}{cc}
   \alpha^2n+(1-n)/2 & 0 \\
  0 & -\alpha^2n+(1+n)/2 \\
\end{array}%
\right) ,\e then \b
C=\frac{1}{2}\sqrt{(1-n+2\alpha^2n)(1+n-2\alpha^2n)} .\e In non
relativistic limit $(n=1)$ we have: $C=2\alpha\sqrt{1-\alpha^2}$
and in ultrarelativistic limit $(n=0)$: $C=1/2$. It is interesting
that for maximally entangled state as $\alpha=1/\sqrt{2}$ for all
values of $n$ concurrence is $1$.

\begin{figure}
\includegraphics[width=4in]{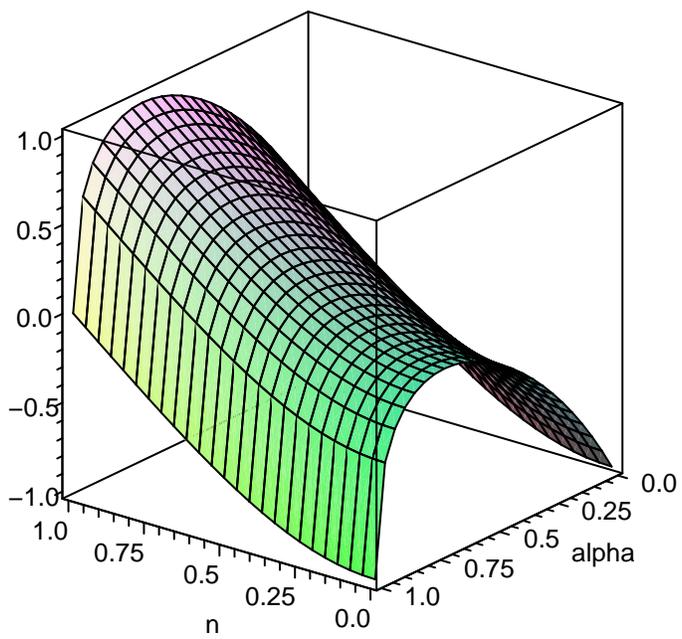}
\newline
\caption{Plot of negativity  versus $n$ and $\alpha$}
 \label{}
\end{figure}

\end{document}